\documentclass[10pt,journal,compsoc]{IEEEtran}

\usepackage[utf8]{inputenc} 
\usepackage[T1]{fontenc}    
\usepackage{hyperref}       
\usepackage{url}            
\usepackage{booktabs}       
\usepackage{amsfonts}       
\usepackage{nicefrac}       
\usepackage{microtype}      
\usepackage{color}
\usepackage{multirow}
\usepackage{graphicx}
\usepackage{pifont}

\begin{document}

\hyphenation{op-tical net-works semi-conduc-tor}

\title{When Large Language Models Meet Personalization: Perspectives of Challenges and Opportunities} 

\author{\IEEEauthorblockN{Jin~Chen\IEEEauthorrefmark{1},
Zheng~Liu\IEEEauthorrefmark{1},
Xu~Huang, 
Chenwang~Wu, Qi~Liu, Gangwei~Jiang, Yuanhao~Pu, Yuxuan~Lei, Xiaolong~Chen, Xingmei~Wang, Defu~Lian and Enhong~Chen}
\IEEEcompsocitemizethanks{
\IEEEcompsocthanksitem \IEEEauthorrefmark{1}Equal Contribution 
\IEEEcompsocthanksitem X. Huang, C. Wu, Q. Liu, G. Jiang, Y. Pu, Y. Lei, X. Chen, X. Wang, D. Lian and E. Chen are with the University of Science and Technology of China. Email: \{xuhuangcs, wcw1996, qiliu67, gwjiang, puyuanhao, lyx180812, chenxiaolong, xingmeiwang\}@mail.ustc.edu.cn and \{liandefu, cheneh\}@ustc.edu.cn.\protect
\IEEEcompsocthanksitem Z. Liu is with Huawei Technologies Ltd. Co. E-mail: zhengliu1026@gmail.com. \protect
\IEEEcompsocthanksitem J. Chen is with the University of Electronic Science and Technology of China. E-mail: chenjin@std.uestc.edu.cn.\protect
}
\thanks{
Corresponding author: Defu Lian (E-mail: liandefu@ustc.edu.cn)}}

\IEEEtitleabstractindextext{%
\begin{abstract}
The advent of large language models marks a revolutionary breakthrough in artificial intelligence. With the unprecedented scale of training and model parameters, the capability of large language models has been dramatically improved, leading to human-like performances in understanding, language synthesizing, and common-sense reasoning, etc. Such a major leap-forward in general AI capacity will fundamentally change the pattern of how personalization is conducted. For one thing, it will reform the way of interaction between humans and personalization systems. Instead of being a passive medium of information filtering, like conventional recommender systems and search engines, large language models present the foundation for active user engagement. On top of such a new foundation, user's requests can be proactively explored, and user's required information can be delivered in a natural, interactable, and explainable way. For another thing, it will also considerably expand the scope of personalization, making it grow from the sole function of collecting personalized information to the compound function of providing personalized services. By leveraging large language models as a general-purpose interface, the personalization systems may compile user's requests into plans, calls the functions of external tools (e.g., search engines, calculators, service APIs, etc.) to execute the plans, and integrate the tools' outputs to complete the end-to-end personalization tasks. Today, large language models are still being rapidly developed, whereas the application in personalization is largely unexplored. Therefore, we consider it to be right the time to review the challenges in personalization and the opportunities to address them with large language models. In particular, we dedicate this perspective paper to the discussion of the following aspects: the development and challenges for the existing personalization system, the newly emerged capabilities of large language models, and the potential ways of making use of large language models for personalization. 
\end{abstract}

}

\maketitle

\IEEEdisplaynontitleabstractindextext

\IEEEpeerreviewmaketitle

\section{Introduction}

The emergence of large language models~\cite{zhao2023survey}, which have demonstrated remarkable progress in understanding human expression, is profoundly impacting the AI community. These models, equipped with vast amounts of data and large-scale neural networks, exhibit impressive capabilities in comprehending human language and generating text that closely resembles our own. Among these abilities are reasoning~\cite{huang2022towards}, few-shot learning~\cite{brown2020language}, and the incorporation of extensive world knowledge within pre-trained models~\cite{zhao2023survey}. This marks a significant breakthrough in the field of artificial intelligence, leading to a revolution in our interactions with machines. Consequently, large language models have become indispensable across various applications, ranging from natural language processing and machine translation to creative content generation and chatbot development.
The introduction of ChatGPT, in particular, has gained significant attention from the human community, prompting reflections on the transformative power of large language models and their potential to push the boundaries of what AI can achieve. This disruptive technology holds the promise of transforming how we interact with and leverage AI in countless domains, opening up new possibilities and opportunities for innovation. As these language models continue to advance and evolve, they are likely to shape the future of artificial intelligence, empowering us to explore uncharted territories and unlock even greater potential in human-machine collaboration.

Personalization, the art of tailoring experiences to individual preferences, stands as an essential and dynamic connection that bridges the gap between humans and machines. In today's technologically-driven world, personalization plays a pivotal role in enhancing user interactions and engagements with a diverse array of digital platforms and services. By adapting to individual preferences, personalization systems empower machines to cater to each user's unique needs, leading to more efficient and enjoyable interactions. Moreover, personalization goes beyond mere content recommendations; it encompasses various facets of user experiences, encompassing user interfaces, communication styles, and more. As artificial intelligence continues to advance, personalization becomes increasingly sophisticated in handling large volumes of interactions and diverse user intents. This calls for the development of more advanced techniques to tackle complex scenarios and provide even more enjoyable and satisfying experiences. The pursuit of improved personalization is driven by the desire to better understand users and cater to their ever-evolving needs. As technology evolves, personalization systems will likely continue to evolve, ultimately creating a future where human-machine interactions are seamlessly integrated into every aspect of our lives, offering personalized and tailored experiences that enrich our daily routines.

Large language models, with their deep and broad capabilities, have the potential to revolutionize personalization systems, transforming the way humans interact and expanding the scope of personalization. the interaction between humans and machines can no longer be simply classified as active and passive, just like traditional search engines and recommendation systems. However, these large language models go beyond simple information filtering and they offer a diverse array of additional functionalities. Specifically, user intent will be actively and comprehensively explored, allowing for more direct and seamless communication between users and systems through natural language. Unlike traditional technologies that relied on abstract and less interpretable ID-based information representation, large language models enable a more profound understanding of users' accurate demands and interests. This deeper comprehension paves the way for higher-quality personalized services, meeting users' needs and preferences in a more refined and effective manner. Moreover, the integration of various tools is greatly enhanced by the capabilities of large language models, significantly broadening the possibilities and scenarios for personalized systems. By transforming user requirements into plans, including understanding, generating, and executing them, users can access a diverse range of information and services. Importantly, users remain unaware of the intricate and complex transformations happening behind the scenes, as they experience a seamless end-to-end model. From this point, the potential of large language models in personal is largely unexplored.

This paper addresses the challenges in personalization and explores the potential solutions using large language models. In the existing related work, LaMP~\cite{salemi2023lamp} introduces a novel benchmark for training and evaluating language models in producing personalized outputs for information retrieval systems. On the other hand, other related surveys~\cite{wu2023survey,lin2023can,fan2023recommender} focus mainly on traditional personalization techniques, such as recommender systems. From the perspective of learning mechanisms, LLM4Rec~\cite{wu2023survey} delves into both Discriminative LLM for Recommendation and Generative LLM for Recommendation. Regarding the adaptation of LLM for recommender systems in terms of 'Where' and 'How', Li et al~\cite{lin2023can} concentrate on the overall pipeline in industrial recommender phases. Fan et al~\cite{fan2023recommender}, on the other hand, conduct a review with a focus on pre-training, fine-tuning, and prompting approaches. While these works discuss pre-trained language models like Bert and GPT for ease of analysis, they dedicate limited attention to the emergent capabilities of large language models. This paper aims to fill this gap by examining the unique and powerful abilities of large language models in the context of personalization, and further expand the scope of personalization with tools.

The remaining of this survey is organized as follows: we review the personalization and large language models in Section 2 to overview the development and challenges. Then we carefully discuss the potential actors of large language models for personalization from Section 3, following the simple utilization of emergent capabilities and the complex integration with other tools. We also discuss the potential challenges when large language models are adapted for personalization.

\section{Background Overview}
\subsection{Personalization Techniques} \label{sec:personalization_sec}
Personalization, a nuanced art that tailors experiences to the unique preferences and needs of individual users, has become a cornerstone of modern artificial intelligence. In this section, we explore the captivating world of personalized techniques and their profound impact on user interactions with AI systems. We will delve into three key aspects of personalization: recommender systems, personalized assistance, and personalized search. These techniques not only enhance user satisfaction but also exemplify the evolution of AI, where machines seamlessly integrate with our lives, understanding us on a profound level. By tailoring recommendations, providing customized assistance, and delivering personalized search results, AI systems have the potential to create a truly immersive and individualized user experience.

\subsubsection{Recommender Systems}
Recommender systems play a pivotal role in personalization, revolutionizing the way users discover and engage with content. These systems aim to predict and suggest items of interest to individual users, such as movies, products, or articles, based on their historical interactions and preferences.

Regarding the development of recommender systems, they have evolved significantly over the years, with collaborative filtering~\cite{resnick1994grouplens,pan2008one} being one of the earliest and most influential approaches. Collaborative filtering relies on user-item interaction data to identify patterns and make recommendations based on users with similar preferences. Traditional solutions, such as matrix factorization~\cite{koren2009matrix} and user/item-based approaches~\cite{wang2006unifying}, extract potentially interesting items based on the idea that users who have shown similar preferences in the past are likely to have similar preferences in the future. While effective, collaborative filtering has limitations, such as the "cold start" problem for new users and items.
To address these limitations, content-based filtering~\cite{pazzani2007content} emerged, which considers the content of items to make recommendations. It leverages the features and attributes of items to find similarities and make personalized suggestions. These features can be grouped into user-side information, such as user profiles, item-side information~\cite{wang2011collaborative,zhang2016collaborative}, such as item brands and item categories, and interaction-based information~\cite{liu2019nrpa}, such as reviews and comments. 
However, content-based filtering may struggle to capture complex user preferences and discover diverse recommendations restricted by the limited feature representations.

In recent years, deep learning has gained significant attention in the field of recommender systems due to its ability to model complex patterns and interactions in user-item data~\cite{wang2015collaborative}. Deep learning-based methods have shown promising results in capturing sequential, temporal, and contextual information, as well as extracting meaningful representations from large-scale data. With the introduction of deep networks, high-order interactions between features of users and items are well captured to extract user interest. Deep learning-based methods offer approaches to capture high-order interactions by employing techniques like attention mechanisms~\cite{zhou2018deep,zhou2019deep} and graph based networks~\cite{wang2019neural} to mining complex relationships between user and item. These methods have been shown to enhance recommendation performance by considering higher-order dependencies and inter-item relationships. Another area of deep learning-based recommender systems is sequential recommenders, specifically designed to handle sequential user-item interactions, such as user behavior sequences over time. Self-Attentions~\cite{kang2018self} and Gated Recurrent Units (GRUs)~\cite{hidasi2015session} are popular choices for modeling sequential data in recommender systems. These models excel in capturing temporal dependencies and context, making them well-suited for tasks like next-item recommendation and session-based recommendation. Sequential-based models can take into account the order in which items are interacted with and learn patterns of user behavior that evolve over time. Furthermore, the rise of language models like BERT has further advanced recommender systems by enabling a better understanding of both natural language features and user sequential behaviors~\cite{sun2019bert4rec}. These language models can capture deep semantic representations and world knowledge, enriching the recommendation process and facilitating more personalized and context-aware recommendations.
Overall, the application of deep learning techniques in recommender systems has opened new avenues for research and innovation, promising to revolutionize the field of personalized recommendations and enhance user experiences.

\subsubsection{Personalized Assistance}
Personalization Assistance refers to the use of artificial intelligence and machine learning techniques to tailor and customize experiences, products, or content based on individual preferences, behavior, and characteristics of users. By analyzing individual preferences, behaviors, and characteristics, it creates a personalized ecosystem that enhances user engagement and satisfaction.
In contrast to traditional recommender systems, which rely on predicting user interests passively, personalized assistance takes a more proactive approach. It ventures into the realm of predicting users' next intentions or actions by utilizing contextual information, such as historical instructions and speech signals. This deeper level of understanding enables the system to cater to users' needs in a more anticipatory and intuitive manner.
At the core of this capability lies the incorporation of cutting-edge technologies like natural language processing (NLP) and computer vision. These advanced tools empower the system to recognize and interpret user intentions, whether conveyed through spoken or written language, or even visual cues.
Moreover, the potential of personalized assistance extends beyond static recommendations to dynamic and context-aware interactions. As the system becomes more familiar with a user's preferences and patterns, it adapts and refines its recommendations in real-time, keeping pace with the ever-changing needs and preferences of the user.

Conversational Recommender Systems mark a remarkable stride forward in the realm of personalized assistance. By engaging users in interactive conversations, these systems delve deeper into their preferences and fine-tune their recommendations accordingly. Leveraging the power of natural language understanding, these conversational recommenders adeptly interpret user queries and responses, culminating in a seamless and engaging user experience.
Notable instances of personalized assistance products, such as Siri and Microsoft Cortana, have already proven their effectiveness on mobile devices. Additionally, the integration of large language models like ChatGPT further elevates the capabilities of conversational recommenders, promising even more enhanced user experiences.
As this technology continues to progress, we can anticipate its growing significance across diverse industries, including healthcare, education, finance, and entertainment. 
While the growth of conversational recommenders and personalized assistance promises immense benefits, it is imperative to develop these products responsibly. Upholding user privacy and ensuring transparent data handling practices are essential to maintain user trust and safeguard sensitive information.

\subsection{Large Language Models}\label{sec:llms}
Language models perform the probabilistic modeling for the generation of natural language, i.e., presented with one specific context, the language models make predictions for the words which are to be generated for the future steps. Nowadays, the language models are mostly built upon deep neural networks, where two features need to be emphasized. First of all, the majority of language models are based on transformers or its close variations \cite{vaswani2017attention}. Such types of neural networks are proficient at modeling context dependency within natural languages, and exhibit superior and consistently improved performances when being scaled up. Secondly, the language models are pre-trained at scale with a massive amount of unlabeled corpus. The pre-trained models are further fine-tuned with task-oriented data so as to adapt to different downstream applications. 

There have been tremendous progresses about language models in recent years, where the emergent of large language models, represented by GPT-3, marks an important milestone for the entire AI community. The large language models (LLMs), as the name suggests, are massively scaled-up derivatives of conventional language models. Particularly, the backbone networks and the training data have been largely magnified. For one thing, although there is no specific criteria for the minimum number, a typical LLM usually consists of no less than several billions and up-to trillions of model parameters, which are orders of larger than before. For another thing, the pre-training is conducted based on much more unsupervised corpora, with hundreds of billions or trillions of tokens carefully filtered from sources like Common Crawl, GitHub, Wikipedia, Books, ArXiv, etc. The impact of scaling is  illustrated by the scaling laws \cite{kaplan2020scaling,hoffmann2022training}, which numerically uncover the power-law relationship between model size, data volume, training scale and the growth of model's performance.

The scaling up of network and training data lead to the leap-forward of large language models' capability. They not only become more proficient at conventional skills, like understanding people's intent and synthesising human-like languages, but also process capabilities which are rarely exhibited by those smaller models. Such a phenomenon is referred as the emergent abilities of LLMs, where three representatives capabilities are frequently discussed. One is the in-context learning capability, where LLMs may quickly learn from the few-shot examples provided in the prompt. Another one is the instruct following capability. After fine-tuned with diversified tasks in the form of instruction tuning, the LLMs are made proficient to follow the human's instructions. Thus, they may handle different tasks presented in an ad-hoc manner. Last but not least, LLMs are found to be able to conduct step-by-step reasoning. With certain types of prompting strategies, like Chain-of-Thought (CoT), LLMs may iteratively approach the final answer of some complex tasks, like mathematical word problems, by breaking down the tasks into sub-problems and figuring out the plausible intermediate answers for each of the sub-problems. 

Thanks to the superior capabilities on understanding, reasoning, and generating, large language models, especially the chat models produced by instruction tuning, are presented as foundamental building blocks for many personalization services. One direct scenario is the conversational search and recommendation. Once built upon large language models, the search and recommendation systems will be able to engage with user via interactions, present outputs in a verbalized and explainable way, receive feedback from the user and make adjustment on top of the feedback, etc. The above changes will bring about a paradigm shift for the personalization services, from passively making search and recommendation, to proactively figuring out user's need and seeking for user's preferred items. In broader scopes, the LLMs may go beyond simply making personalized search and recommendation, but play as personalized assistants to help users with their task completions. The LLMs may take notes of users' important information within their memory, make personalized plans based on memorized information when new demands are raised, and execute plans by leveraging tools like search engines and recommendation systems. 

Yet, we have to confront the reality that applying LLMs for personalization is not a trivial problem. To name a quite few of the open challenges. Firstly, personalization calls for the understanding of user preference, which is more of domain-specific knowledge rather than the common-sense knowledge learned by LLMs. The effectively and efficiently adaptation of LLMs for personalized services remains to be resolved. Besides, the LLMs could memorize user's confidential information while providing personalized services. Thus, it raises the concerns for privacy protection. The LLMs are learned from Internet data; due to the exposure bias, it is almost inevitable to make unfair  predictions for the minorities. To address the above challenges, benchmarks and evaluation datasets are needed by the research communities. However, such resources are far from complete at present. To fully support personalization with LLMs, methodological and experimental frameworks need to be systematically established for all these perspectives. 

\section{LLMs for Personalization}
\begin{figure*}
    \centering
    \includegraphics[width=0.96\textwidth]{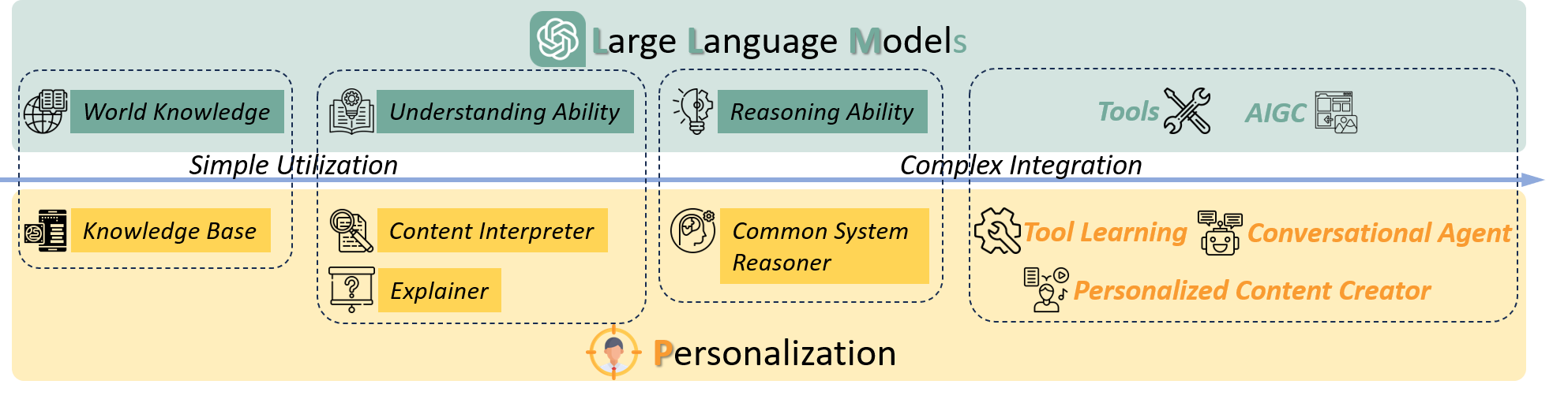}
    \caption{The Overview of LLM for Personalization}
    \label{fig:overall}
\end{figure*}

In the following sections, we delve into the potential of large language models for personalization, examining their evolution from simple use cases, like utilizing word knowledge as features, to more intricate integration with other tool modules to act as agents. Specifically, we focus on the progression of emergent capabilities, starting from basic world knowledge and understanding user intent, and advancing to high-level reasoning abilities. We explore how large language models can contribute to constructing a knowledge base that enriches common-sense knowledge about various items. Additionally, we discuss how the understanding capability of large language models can empower content interpreters and explainers for in-depth analysis of interactions.
Furthermore, we observe attempts to leverage the reasoning ability of large language models for system reasoners to provide recommendation results. 
These increasingly sophisticated capabilities enable complex utilization of large language models with other tool modules, enabling them to better comprehend user intentions and fulfill user instructions. Consequently, we also explore the integration of large language models with other tools for personalization, including tool learning, conversational agents and personalized content creators. The overview of this chapter is depicted in Figure~\ref{fig:overall}. Our comprehensive survey aims to provide a deeper understanding of the current landscape, shedding light on the opportunities and challenges associated with incorporating large language models into personalization.

\section{LLMs as Knowledge Base}\label{sec:llm_knowledge_base} 


Knowledge base provides rich information with semantics, attracting increasing attention for the usage of the knowledge base in the recommender systems. Particularly, the knowledge graphs, where nodes represent entities and edges represent relations in the heterogeneous information graph, are the common format of knowledge bases and introduced as side information to enhance the performance of recommenders. Knowledge graphs help understand the mutual relations between users and items and also provides better explainability for recommenders. Existing methods that incorporate knowledge graphs in recommender systems can be classified into three main groups: embedding-based methods, path-based methods and the unified methods. Embedding-based methods, such as CKE~\cite{zhang2016collaborative} and DKN~\cite{wang2018dkn}, KSR~\cite{huang2018improving}, SHINE~\cite{wang2018shine}, utilize semantic representations of users and items. These methods aim to capture the underlying semantic relationships between entities in the knowledge graph, which can improve the quality of recommendations. Path-based approaches, such as Hete-MF~\cite{yu2013collaborative}, SemRec~\cite{shi2015semantic}, RuleRec~\cite{ma2019jointly}, EIUM~\cite{huang2019explainable},exploit the semantic connectivity information present in the knowledge graph to regularize the user and item representations. These methods consider the paths between users and items in the graph and leverage them to incorporate interpretability into the recommendation process. Unified methods, such as RippleNet~\cite{wang2018ripplenet}, KGCN~\cite{wang2019knowledge}, KGAT~\cite{wang2019kgat}, AKUPM~\cite{tang2019akupm}, IntentGC~\cite{zhao2019intentgc} refine the representations of entities in the knowledge graph by leveraging embedding propagation techniques. These methods propagate the embeddings of entities through the graph structure, allowing information to flow across connected entities and refining the representations accordingly.

However, the knowledge graphs adopted in recommender systems is limited and with low usability. Reviewing the various knowledge graph datasets for recommender systems, covering the domains of movie, book, news, product, etc., these datasets are still significantly sparse compared to the vast amount of human knowledge, particularly the lack of facts, due to the expensive supervision to construct the knowledge graph. Building a comprehensive and accurate knowledge graph would be a complex and resource-intensive task, which would include data collection, integration, and cleaning to assure data quality and consistency. Limited by the expensive cost of labelling the knowledge graphs, there would usually exist missing entities or relations. The user preferences for these entities or paths may be ignored, and the recommendation performance suffers.

The ability of Large Language Models to retrieve factual knowledge as explicit knowledge bases~\cite{petroni2019language,roberts2020much,petronicontext,jiang2020can,wang2020language,poerner2020bert,petronicontext,jiang2020can,heinzerling2021language,wang2021can,guu2020retrieval} has been stirred discussed, which presents an opportunity to construct more comprehensive knowledge graphs within recommender systems. Tracing back to the work~\cite{petroni2019language}, large language models have shown their impressive power in storing factual information, such as entities and common-sense, and then commonsense knowledge can be reliably transferred to downtown tasks. 
\textbf{Existing methods in knowledge graphs fall short of handling incomplete KGs~\cite{bordes2013translating} and constructing KGs with text corpus~\cite{zhu2023llms}} and many researchers attempt to leverage the power of LLM to solve the two tasks, i.e., the knowledge completion~\cite{zhang2020pretrain} and knowledge construction~\cite{kumar2020building}. \textbf{For knowledge graph completion}, which refers to the task of missing facts in the given knowledge graph, recent efforts have been paid to encode text or generate facts for knowledge graphs. MTL-KGC~\cite{kim2020multi} encoders the text sequences to predict the possibility of the tuples. MEMKGC~\cite{choi2021mem} predicts the masked entities of the triple. StAR~\cite{wang2021structure} utilizes Siamese textual encoders to separately encode the entities. GenKGC~\cite{xie2022discrimination} uses the decoder-only language models to directly generate the tail entity. TagReal~\cite{jiang2023text} generates high-quality prompts from the external text corpora. AutoKG~\cite{zhu2023llms} directly adopts the LLMs, such as ChatGPT and GPT-4, and design tailored prompts to predict the tail entity.
As for the another important task, i.e., \textbf{knowledge graph construction}, which refers to creating a structured representation of knowledge, LLMs can be applied in the process of constructing knowledge graphs, including entity discovery~\cite{yan2021unified,li2022ultra}, coreference resolution~\cite{kirstain2021coreference,cattan2021cross} and relation extraction~\cite{lyu2021relation,wang2019fine}. LLMs can also achieve the end-to-end construction~\cite{han2023pive,kumar2020building,wang2020language,trajanoska2023enhancing,jiang2023text} to directly build KGs from raw text. 
LLMs enables the knowledge distillation to construct knowledge graphs. symbolic-kg~\cite{west2022symbolic} distills commonsense facts from GPT3 and then finetune the small student model to generate knowledge graphs.
These models have demonstrated the capacity to store large volumes of knowledge, providing a viable option for improving the scope and depth of knowledge graphs. Furthermore, these advancements have prompted research into the direct transfer of stored knowledge from LLMs to knowledge graphs, eliminating the need for human supervision. This interesting research throws light on the possibilities of automating knowledge graph completion utilizing cutting-edge big language models.

By leveraging the capabilities of LLMs, recommender systems would benefit from a more extensive and up-to-date knowledge base. Firstly, missing faculty information can be completed to construct more extensive knowledge graphs and thus the relations between entities can be extracted for better recommenders. Secondly, in contrast to the preceding exclusively in-domain data, the large language model itself contains plenty of cross-domain information that can help achieve cross-domain recommendations, such as recommending appropriate movies based on the user's favorite music songs. To sum up, the stored knowledge can be utilized to enhance recommendation accuracy, relevance, and personalization, ultimately improving the overall performance of recommender systems. Existing work~\cite{xi2023towards} prompts the large language models to generate the factual knowledge about movies to enhance the performance of CTR prediction models. To better utilize the factual knowledge, a \textit{Knowledge Adaptation} module is adopted for better contextual information extraction.  

It is worth noting that the \textbf{phantom} problem of large language models can be a challenge when applied to recommendation tasks. The inherent nature of large language models can introduce ambiguity or inaccurate provenance~\cite{razniewski2021language}. This issue can emerge as the introduction of extraneous information or even noise into the recommendation process. The large language models may generate responses that, while syntactically correct, lack informative context or relevance. 
According to the KoLA~\cite{yu2023kola}, a benchmark for evaluating word knowledge of LLMs,  even the top-ranked GPT4 just achieves 0.012 in Precision and 0.013 in Recall on the task \textit{Named Entity Recognition}, which falls far short of the performance (0.712 in Precision and 0.706 in Recall) of the task specific models PL-Marker~\cite{ye2022packed}. Such a finding suggests that common sense is still far from being sufficiently captured by LLM.
By aggregating the results with irrelevant or deceptive information, this can damage the usefulness of the recommendation system.

\section{LLMs as Content Interpreter}
Content-based recommenders provide an effective solution for mitigating the sparse feedback issue in recommender systems. By leveraging the attributes and characteristics of items, these systems achieve a more profound understanding of their properties, facilitating accurate matching with user preferences. However, the content features used in content-based recommendation may also exhibit sparsity. Relying solely on the recommended supervision signal, such as clicking and browsing, might not fully exploit the potential benefits of these features. To overcome this challenge, language models emerge as powerful fundamental algorithms that act as content interpreters in processing textual features. Their utilization enhances the effectiveness of recommender systems by effectively understanding and interpreting textual content, leading to improved recommendations. 

\begin{figure*}
    \centering
    \includegraphics[width=1\textwidth]{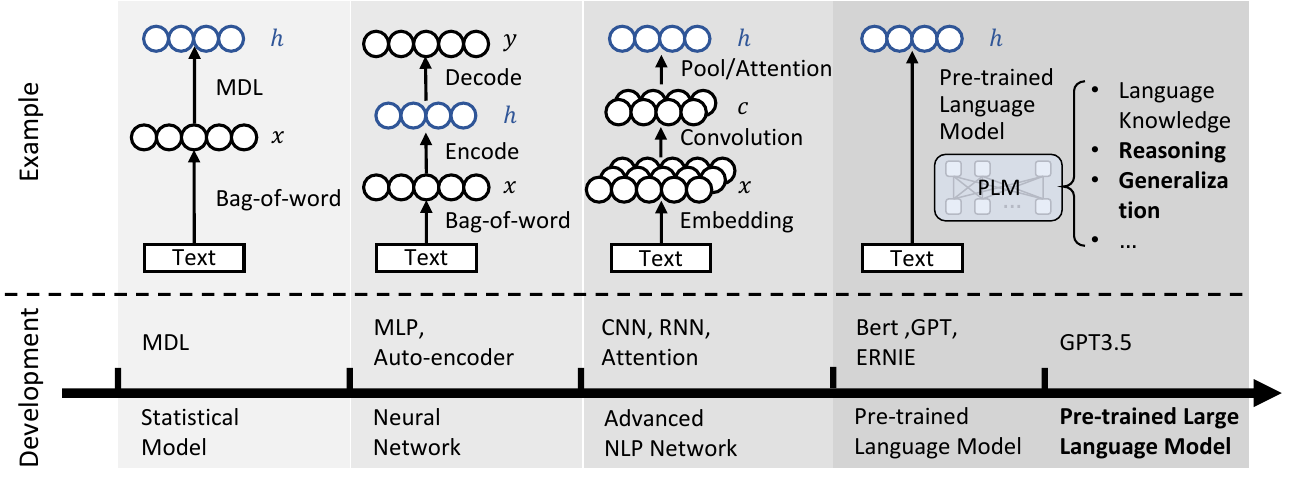}
    \caption{The development of content interpreter in recommendation.}
    \label{fig:content_interpreter}
\end{figure*}

\subsection{Conventional Content Interpreter}
Conventional content interpreter includes statistical model, neural network, and advanced NLP network, as summarized in Figure~\ref{fig:content_interpreter}. These approaches primarily focus on transforming content information, such as textual data, into feature embeddings to facilitate the recommendation process.

Statistical models like TF-IDF, Minimum Description Length (MDL)~\cite{lang1995newsweeder}, and bag-of-words have been traditionally used to encode textual data such as news articles and documents into continuous value vectors. However, with the advancement of deep learning techniques, researchers have explored various neural network architectures to learn more expressive content representations. Instead of relying solely on statistical embeddings, some approaches initialize the vectors with bag-of-words representations and then employ autoencoder-based models to learn more powerful representations.
For example, CDL~\cite{wang2015collaborative} combines the latent vectors obtained from autoencoders with the original ID embeddings to enhance content representations. CRAE~\cite{wang2016collaborative} introduces a collaborative recurrent autoencoder that captures the word order in texts, enabling the modeling of content sequences in collaborative filtering scenarios. Dong et al. ~\cite{dong2017hybrid} propose a stacked denoising autoencoder that reconstructs item/user ratings and textual information simultaneously, allowing for the joint modeling of collaborative and textual knowledge.
CVAE~\cite{li2017collaborative} introduces a collaborative variational autoencoder that learns probabilistic textual features. While autoencoders are effective in learning low-dimensional representations from text data, they may struggle to capture semantic information effectively~\cite{WuWHX23survey}. In some cases, approaches like doc2vec~\cite{LeM14doc2vec} are used to construct content embeddings~\cite{SongEH16mlp, KumarKG17mlp} and learn hidden representations. Okura et al.~\cite{okura2017embedding} evaluate different network architectures, including word-models and GRU networks, for representing user states.
  
Following the advancements in neural natural language processing (NLP) models, more sophisticated architectures such as Convolutional Neural Networks (CNNs), Recurrent Neural Networks (RNNs), and Neural Attention models have been employed as content interpreters to extract contextual information and capture user preferences. These models take sentence inputs, such as news titles, reviews, or comments, and transform them into word embedding matrices using random initialization or word2vec embeddings~\cite{Mikolov2013word2vec}. Various architectures, including CNNs, attention networks, and hybrid models, are utilized to learn representations of sentences.
For example, NPA~\cite{wu2019cnnattn} and LSTUR~\cite{An19cnnrnn} incorporate attention mechanisms to determine the importance of words after CNN layers. NRMS~\cite{wu19selfattn} and CPRS~\cite{wu20selfattn} utilize multi-head self-attention networks to learn word representations. These models are effective in capturing long-context dependencies and understanding the semantic information in the text.
In addition to text modeling, language models are also used as content interpreters to capture user interests based on their historical interactions. For instance, WE3CN~\cite{khattar2018cnn} employs a 3D CNN to extract temporal features from the historical data. DKN~\cite{wang2018dkn} utilizes an attention mechanism to aggregate historical information related to candidate items. DAN~\cite{Zhu19DAN} proposes an attention-based LSTM to capture richer hidden sequence features. These models leverage different neural architectures to enhance the representation of text in the context of recommendation systems.
It is worth noting that these models still have limitations in terms of depth and the ability to effectively generalize semantic information.

\subsection{Language Model based Content Interpreter}
In recent years, there has been a growing interest in incorporating more powerful pre-trained language models, such as BERT and GPT, into recommendation systems. These language models have shown exceptional performance in various natural language processing tasks and have sparked researchers' inspiration to leverage them for capturing deep semantic representations and incorporating world knowledge in recommendation systems.
However, applying pre-trained language models to recommendation tasks presents two main challenges. Firstly, there is a misalignment of goals between general-purpose language models and the specific objectives of recommendation systems. To address this, researchers have proposed approaches that fine-tune the pre-trained models or design task-specific pre-training tasks to adapt them to recommendation tasks. For example, U-BERT~\cite{qiu2021u} employs BERT as a content interpreter and introduces masked opinion token prediction and opinion rating prediction as pre-training tasks to better align BERT with recommendation objectives. Similarly, other works~\cite{zhang2021unbert, wu2021empowering, liu2022boosting, wu2022mm, yu2022tiny} have utilized pre-trained BERT to initialize the news encoder for news recommendation, enhancing the representation of textual features. The pre-trained model, ERNIE, is also utilized to enhance the representation ability of queries and documents~\cite{zou2021pre, liu2021pre}.
The second challenge is reducing the online inference latency caused by pre-trained language models, which can be computationally expensive. Researchers have explored techniques such as knowledge distillation and model optimization to obtain lightweight and efficient models suitable for online services. For instance, CTR-BERT~\cite{muhamed2021ctr} employs knowledge distillation to obtain a cache-friendly model for click-through rate prediction, addressing the latency issue.

Moreover, pre-trained language models have been applied beyond mainstream recommendation tasks. They have been integrated into various recommendation scenarios, including tag recommendation~\cite{he2022ptm4tag}, tweet representations~\cite{zhang2022twhin}, and code example recommendation~\cite{rahmani2023improving}, to enhance the representation of textual features in those specific domains. Additionally, some recent works~\cite{ding2021zero, hou2022towards, hou2023learning, yuan2023go} have explored using only textual features as inputs to recommendation models, leveraging pre-trained language models to alleviate cold-start problems and enable cross-domain recommendations. This paradigm offers advantages in alleviating cold-start problems and facilitating cross-domain recommendations based on the universality of natural language. ZESREC~\cite{ding2021zero} uses BERT to obtain universal continuous representations of item descriptions for zero-shot recommendation. Unisrec~\cite{hou2022towards} focuses on cross-domain sequential recommendation and employs a lightweight MoE-enhanced module to incorporate the fixed BERT representation into the recommendation task. VQ-Rec~\cite{hou2023learning} further aligns the textual embeddings produced by pre-trained language models to the recommendation task with the help of vector quantization. Fu et al.~\cite{fu2023exploring} explore layerwise adaptor tuning to achieve parameter-efficient transferable recommendations.

\begin{table*}[th]
\newcommand{\tabincell}[2]{
\begin{tabular}{@{}#1@{}}#2\end{tabular}
}
    \centering
    \caption{LLMs for Content Interpreter}
    \label{tab:my_label}

    \resizebox{1.\textwidth}{!}{
\begin{tabular}{c|c|c|c|c}
\toprule
Approach & Task & LLM backbone  & Tuning Strategy  & Datasets    \\ \midrule
TALLRe~\cite{bao2023tallrec} & Sequential Recommendation &  LLaMA-7B & Instruct Tuning \& Fine Tuning & MovieLens100k, BookCrossing \\ \midrule
LLMs-Rec~\cite{kang2023llms}& Rating Prediction & Flan-T5-Base, Flan-T5-XXL & Fine Tuning & MovieLens-1M, Amazon Book \\ \midrule
PALR~\cite{chen2023palr} & Item Recommendation & LLaMa-7B &  Instruction Tuning & MovieLens-1M, Amazon Beauty \\ \midrule
InstructRec~\cite{zhang2023recommendation} &  \tabincell{c}{sequential recommendation \\ personalized search} &Flan-T5-XL & Instruction Tuning & Amazon-Games, CDs \\ \bottomrule
\end{tabular}
}
\end{table*}

While the pre-trained language models empower the text understanding with the benefit of capturing world knowledge first, the development of pre-trained large language model provides great emergency ability in the fields of reasoning and generalization. 
TALLRe~\cite{bao2023tallrec} explores the ability of large language models for the  sequential recommendation. They observe that original language models perform poorly in zero-shot and few-shot scenarios, while recommendation-specific instruction-tuned language models demonstrate superior performance in few-shot learning and cross-domain generalization. Similarly, Kang et al.~\cite{kang2023llms} propose a similar instruction tuning method for rating prediction recommendation tasks based on the T5 backbone. They find that the tuned language models, which leverage data efficiency, outperform traditional recommenders. PALR~\cite{chen2023palr} further enhances the construction pipeline of recommendation-specific instruction tuning, which first employs large language models to generate reasoning as additional features based on the user's behavior history. Next, a small set of candidates is retrieved using any existing model based on the user profile. Finally, to adapt general-purpose language models to the recommendation task, they convert the generated reasoning features, user interaction history, and retrieved candidates into natural language instruction data and fine-tune a language model.
Existing instruction tuning methods of language models for recommendation scenarios typically focus on a single type of recommendation task, limiting the full utilization of language models' strong generalization ability. InstructRec~\cite{zhang2023recommendation} addresses this limitation by formulating recommendation as an instruction-following procedure. They design various instruction templates to accommodate different recommendation tasks and employ GPT-3.5 to generate high-quality instruction data based on the user's historical data and templates. The language models fine-tuned using this instruction data can effectively handle a wide range of recommendation tasks and cater to users' diverse information requirements.

\section{LLMs as Explainer}
In addition to valuing the suggestions made by a recommendation model, users are also interested in the comprehensible justifications for these recommendations~\cite{wang2018reinforcement,gao2019explainable}. This is crucial as most recommender systems are black boxes whose inner workings are inscrutable to human understanding~\cite{leeself}, diminishing user trust. Taking drug recommendations, for instance, it is unacceptable to recommend drugs with good curative effects simply but fail to give reasons why they are effective. To this end, explainable recommendations aim to couple high-quality suggestions with accessible explanations. This not only helps to improve the model's transparency, persuasiveness, and reliability, but also facilitates the identification and rectification of potential errors through insightful explanations. These benefits have been extensively documented in recent work~\cite{nye2021show,lampinen2022can,wei2022chain,zelikman2022star}. For instance, ~\cite{lampinen2022can} conducted a study that involved addressing 40 difficult tasks and evaluating the impact of explanations on zero-shot and few-shot scenarios. Their findings demonstrated that explanations have a positive effect on model performance by establishing a connection between examples and interpretation.

Traditional approaches mainly focus on template-based explanations, which can be broadly categorized into item-based, user-based, and attribute-based explanations\cite{zhang2020explainable}. Item-based explainable methods relate recommendations to familiar items~\cite{schafer1999recommender}, explaining that \textit{the recommended item bears similarity to others the user prefers}, which are prevalent on platforms like Amazon~\cite{linden2003amazon} and Netflix~\cite{gomez2015netflix}. However, due to its collaboration, it may underperform in personalized recommendations requiring diversity and can struggle to identify relevant items among industrial settings with vast items efficiently. In contrast, user-based explanations~\cite{sinha2002role} leverage social relationships to make recommendations by explaining that \textit{users with similar interests also favor the recommended item}. The user's social property makes these explanations more persuasive, encouraging users to try the recommendations. However, the variance in user preferences may render this approach less impactful in gauging actual preference. Lastly, attribute-based explanations focus on highlighting the attributes of recommended items that users might find appealing, essentially conveying "\textit{these features might interest you}". This method demands customization according to each user's interests, yielding higher accuracy and satisfaction. Thus, they are at the forefront of research~\cite{wang2018reinforcement, xian2021ex3, wang2022reinforced, verma2022recxplainer,zhang2022neuro}.

Obviously, such explanations typically employ pre-defined and formulaic formats, such as explanations based on similar items or friends. Although capable of conveying essential information, such inflexible formats may diminish the user experience and satisfaction by lacking adaptability and personalization~\cite{wang2018reinforcement}. For this reason, natural language generation approaches have received increasing attention. Early work~\cite{li2017neural,dong2017learning,li2020generate} mainly relied on recurrent neural networks (e.g., LSTM~\cite{hochreiter1997long}, GRU~\cite{cho2014learning}). Limited by the model's expressiveness, they often suffer from the issue of insufficient diversity. With the excellent performance of Transformer-based models in various natural language tasks, some work attempts to integrate Transformer-based models into explainable recommendations. ~\cite{li2021personalized} use the position vectors corresponding to the user (item) IDs to predict interpreted tokens. Subsequent work~\cite{zhan2023towards} has shown that the generated explanation cannot justify the user's preference by synthesizing irrelevant descriptions. Therefore, Ni et al.~\cite{ni2019justifying} used such information as guided input to BERT to obtain a controllable justification. Considering that such auxiliary information is not always available in real-world scenarios, ExBERT~\cite{zhan2023towards} only requires historical explanations written by users, and utilizes a multi-head self-attention based encoder to capture the relevance between these explanations and user-item pairs. Recently, MMCT~\cite{liu2023multimodal}, EG4Rec~\cite{qu2022explanation}, and KAER~\cite{bai2020fusing} have further carried out finer-grained modeling of information such as visual images, time series, and emotional tendencies to obtain high-quality interpretations.

Due to the limited expressive power of traditional language models, natural language generation methods are prone to long-range dependence problems~\cite{zhan2023towards}, that is, the input of long texts will appear to generate explanations that lack diversity and coherence in content. In addition, these explanation methods are tightly coupled with specific recommendation models (e.g., NETE~\cite{li2020generate}), or directly design a new recommendation model (e.g., NRT~\cite{li2017neural}, PETER~\cite{li2021personalized}), and they are often powerless when faced with existing advanced recommendation models, which limits their generalizability. This is also a flaw in template-based methods. Notably, in industrial settings, recommendation algorithms frequently involve not just a single model but a cascade or integration of multiple models, and these elaborate combinations further exacerbate the difficulty of deciphering recommendations.

Thanks to LLMs' remarkable generative ability in language tasks, making them ideal for tackling the aforementioned challenges~\cite{bommasani2021opportunities}. Firstly, with the leverage of extensive training data, LLMs adeptly harness human language, encompassing context, metaphors, and complex syntax. This equips them to craft customized explanations that are precise, natural, and adaptable to various user preferences~\cite{li2020generate,li2021personalized,li2023personalized}, mitigating the limitations of conventional, formulaic explanations. 
Secondly, the unique in-context learning capabilities of LLMs, such as zero-shot prompting, few-shot prompting, and chain-of-thought prompting, enable them to garner real-time user feedback during interactions, furnish recommendation outcomes, and their corresponding interpretations, fostering bidirectional human-machine alignment. Recent study~\cite{bills2023language} has demonstrated the potential of LLMs in elucidating the intricacies of complex models, as evidenced by GPT-4 autonomously interpreting the function of GPT-2's each neuron by inputting appropriate prompts and the corresponding neuron activation. This showcases an innovative approach to interpreting deep learning-based recommendation models. It’s critical to highlight that this interpretation technique is agnostic to the model's architecture, distinguishing it from traditional interpretations that are bound to specific algorithms. Thus, recommendation interpretations founded on LLMs pave the way for a versatile and scalable interpretational framework with broader applicability.

Although LLMs have inherently significant advantages in recommendation explanations, it is imperative to recognize potential issues. Firstly, akin to recommendation models, LLMs are essentially black boxes that are difficult for humans to understand. We cannot identify what concepts they give explanations based on~\cite{wu2023interpretability}. Also, the explanation given may be insincere; that is, the explanations are inconsistent with their recommended behaviors. Some recent developments~\cite{li2023making,wei2022chain} involve utilizing chains of thought to prompt reasoning for improved interpretability; however, the opacity of the reasoning process of each step remains a concern, and ~\cite{turpin2023language} has questioned the possible unfaithful explanations of chain-of-thought prompting. Secondly, the extensive data utilized by LLMs may encompass human biases and erroneous content~\cite{li2021hidden}. Consequently, even if the explanation aligns with the model’s recommendation behavior, both the explanation and recommendation could be flawed. Monitoring and calibrating these models to ensure fairness and accuracy in explainable recommendations is essential. Lastly, generative models exhibit varying levels of proficiency across different tasks, leading to inconsistencies in performance. Identical semantic cues could yield disparate recommendation explanations. This inconsistency has been substantiated by recent studies~\cite{wang2023robustness,han2023information} focusing on the LLMs' robustness. Addressing these issues calls for exploring techniques to mitigate or even circumvent low-reliability explanatory behavior, and investigating how LLMs can be trained to consistently generate reliable recommendation explanations, especially under adversarial conditions, is a worthwhile avenue for further research.

\section{LLMs as Common System Reasoner}
With the development of large language models, there is an observation that LLMs exhibit reasoning abilities~\cite{huang2022towards,wei2022emergent} when they are sufficiently large, which is fundamental for human intelligence for decision-making and problem-solving. By providing the models with the `chain of thoughts'~\cite{wei2022chain}, such as prompting with \textit{'let us think about it step by step'}, the large language models exhibit emergent abilities for reasoning and can arrive at conclusions or judgments according to the evidence or logics. Accordingly, for recommender systems, large language models are capable of reasoning to help user interest mining, thus improving performance. 
 
\subsection{Making Direct Recommendations}

\begin{table*}[]
\newcommand{\tabincell}[2]{
\begin{tabular}{@{}#1@{}}#2\end{tabular}
}

    \caption{Zero/few-shot learners of LLMs for RS}
    \label{tab:zero_show_recs}
    \resizebox{1.\textwidth}{!}{
\begin{tabular}{c|c|c|c|c|c|c}
\toprule
Approach                                                                & LLM backbone                                                                & Task                           & Metric                    & Datasets    & ICL & COT                                                      \\ \midrule
\cite{liu2023chatgpt}    & gpt-3.5-turbo                                   & \tabincell{c}{rating prediction \\ sequential recommendation \\  direct recommendation \\   explanation generation \\ review summarization }         & \tabincell{c}{RMSE,MAE \\HR,NDCG \\ HR,NDCG \\     BLUE4,ROUGE,Human Eval\\BLUE4,ROUGE,Human Eval }             & Amazon Beauty   & $\checkmark$ &                                  \\ \midrule
\cite{dai2023uncovering} & \tabincell{c}{text-davinci-002\\text-davinci-003\\gpt-3.5-turbo}  & \tabincell{c}{point-wise\\pair-wise\\list-wise}                     & NDCG,MRR & \tabincell{c}{MovieLens-1M\\Amazon-Book\\Amazon-Music\\MIND-small} & $\checkmark$ & \\  \midrule
\cite{kang2023llms}            & \tabincell{c}{Flan-U-PALM\\gpt-3.5-turbo \\text-davinci-003} & \tabincell{c}{rating prediction\\ranking prediction}              & \tabincell{c}{RMSE,MAE \\ROC-AUC}                  & \tabincell{c}{MovieLens-1M\\Amazon-Books}                 & $\checkmark$ & \\ \midrule
\cite{wang2023zero}                              & text-davinci-003                                                     & reranking& NDCG,HR                   & MovieLens 100K  & $\checkmark$&$\checkmark$                                                  \\ \midrule
\cite{hou2023large}                              & gpt-3.5-turbo                                                               & reranking& NDCG                      & \tabincell{c}{MovieLens-1M\\Amazon-Games}   & $\checkmark$ &   \\ \midrule
~\cite{li2023preliminary} & gpt-3.5-turbo & reranking & Precision & MIND & $\checkmark$ & \\ \bottomrule                                   
\end{tabular}
}
\end{table*}

In-context learning~\cite{dong2022survey,dai2022can,min2022rethinking,levy2022diverse,xieexplanation,olsson2022context,akyurek2022learning} is one of the emergent abilities of LLMs that differentiate LLMs from previous pre-trained language models, where, given a natural language instruction and task demonstrations, LLMs would generate the output by completing the word sequence without training or tuning~\cite{brown2020language}. As for in-context learning, the prompt follows by the task instruction and/or the several input-output pairs to demonstrate the task and a test input is added to require the LLM to make predictions. The input-output pair is called a \textit{shot}. This emergent ability enables prediction on new cases without tuning unlike previous machine learning.

\begin{figure*}
    \centering
    \includegraphics[width=0.95\textwidth]{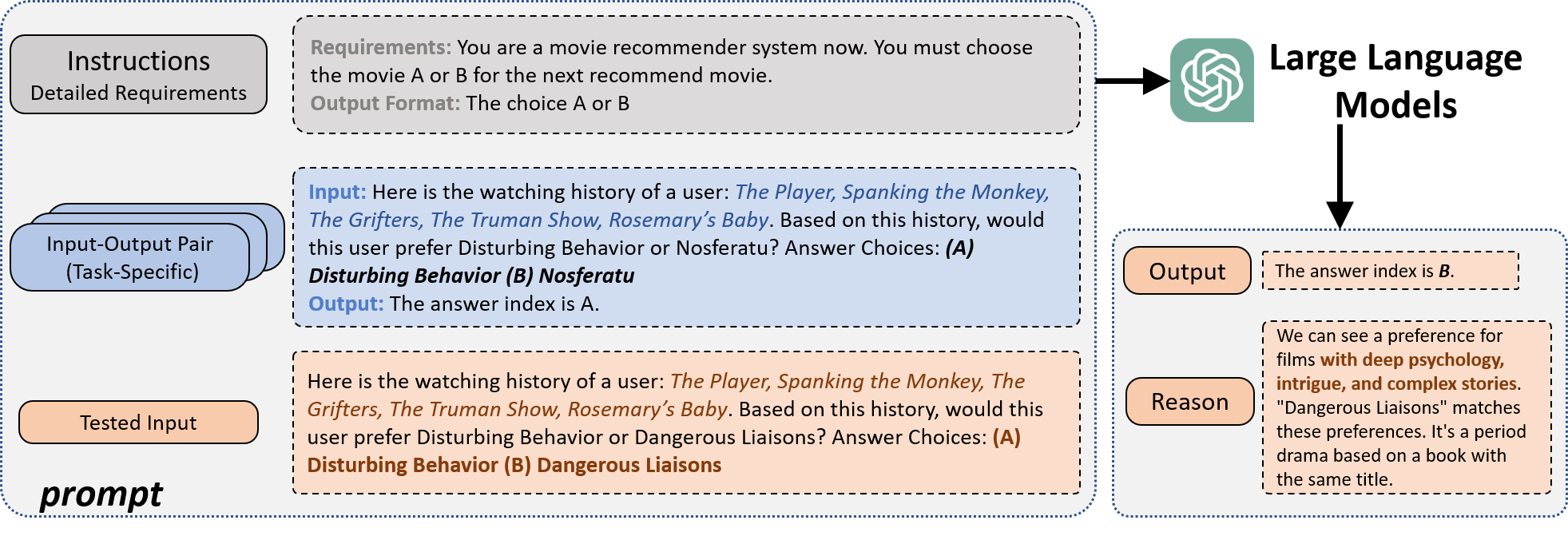}
    \caption{An Example of zero/few-shot learning for direct recommenders}
    \label{fig:zero_few_shot}
\end{figure*}

In the realm of recommender systems, numerous studies have explored the performance of zero-shot/few-shot learning using large language models, covering the common recommendation tasks such as rating prediction, and ranking prediction. These studies evaluate the ability of language models to provide recommendations without explicit tuning, as summarized in Table~\ref{tab:zero_show_recs}, where all methods adopt in-context learning for direct recommenders. The general process can be attached in Figure~\ref{fig:zero_few_shot}. Accordingly, we have the following findings:
\begin{itemize}
    \item The aforementioned studies primarily focused on evaluating zero-shot/few-shot recommenders using open-domain datasets, predominantly in domains such as movies and books. Large language models are trained on extensive open-domain datasets, enabling them to possess a significant amount of common-sense knowledge, including information about well-known movies. However, when it comes to private domain data, such as e-commerce products or specific locations, the ability of zero-shot recommenders lacks of validation, which is expected to be challenging.
    \item Current testing methods necessitate the integration of additional modules to validate the performance of zero-shot recommenders for specific tasks. In particular, for ranking tasks that involve providing a list of items in order of preference, a candidate generation module is employed to narrow down the pool of items~\cite{wang2023zero} and \cite{hou2023large}. Generative-based models like gpt-3.5-turbo generate results in a generative manner rather than relying on recall from existing memories, thus requiring additional modules to implement ID-based item recommendations.
    \item From the perspective of recommendation performance, zero-shot recommenders exhibit some capabilities and few-shot learners perform better than zero-shot recommenders. However, there still exists a substantial gap when compared to traditional recommendation models, particularly fine-tuned large language models designed specifically for recommenders, such as P5\cite{geng2022recommendation} and M6-Rec~\cite{cui2022m6}. This highlights that large language models do not possess a significant advantage in personalized modeling.
\end{itemize}

Another important emergent ability is the \textit{`step by step'} reasoning, where LLMs can solve complex tasks by utilizing prompts including previous intermediate reasoning steps, called the `chain of thoughts' strategy~\cite{wei2022chain}. Wang and Lim~\cite{wang2023zero} design a three-step prompt, namely NIR, to capture user preferences, extract the most representative movies and rerank the items after item filtering. Such a multi-step reasoning strategy significantly improves recommendation performance.

\subsection{Reasoning for Automated Selection}
Automated Machine Learning (AutoML) is widely applied in recommender systems to eliminate the costly manual setup with trials and errors. The search space in recommender systems can be categorized in (1) Embedding size (2) Feature (3) Feature interaction (4) Model architecture. Embedding size search, such as~\cite{liu2021learnable,liu2020automated,deng2021deeplight,ginart2021mixed} seeks for appropriate embedding size for each feature to avoid resources overconsumption. Searching for features consisting of raw feature search\cite{wang2022autofield,lin2022adafs} and synthetic feature search\cite{tsang2020feature,yuanfei2019autocross}, which selects a subset from the set of original or cross features to maintain informative features to reduce both computation and space cost. Feature interaction search, such as~\cite{liu2020autofis,liu2020autogroup, chen2019bayesian,xie2021fives,su2021detecting}, automatically filters out feature interactions that are not helpful. Model architecture search, like~\cite{song2020autoctr,zhao2021ameir,wei2021autoias,cheng2022nasr}, expands the search space to the integral architectures. The search strategy shifts from the discrete reinforcement learning process, which iteratively samples architectures for training and is time-consuming, into the differentiable searching, which adaptively selects architectures within one-shot learning to circumvent the computational burden, for more efficient convergence. The evaluation for each sampled architecture then acts as the signal to adjust the selections. That is, there is a decision maker who memorizes the prior results of previous architecture choices and analyzes the prior results to give the next recommended choice.

The emergent LLMs actually have excellent memorization and reasoning capability that would work for automated learning. Several works have attempted to validate the potential of automated machine learning with LLMs. Preliminarily, GPT-NAS~\cite{yu2023gptnas} takes advantage of generative capability of LLMs. The architecture of networks are formulated into sequential characters, and thus the generation of network architectures can be easily achieved through the generative pre-training models. NAS-Bench-101~\cite{ying2019bench} is utilized for pre-training and the state-of-the-art results are used for fine-tuning. The generative pre-training models produce reasonable architectures, which would reduce the search space for later genetic algorithms for searching optimal architectures. The relatively advanced reasoning ability is further evaluated in GENIUS~\cite{zheng2023cangpt}, where GPT-4 is employed as a black-box agent to generate potential better-performing architectures according to previous trials including tried architectures with their evaluation performance. According to the results, GPT-4 can generate good architecture networks, showing the potential for more complicated tasks. Yet it is too difficult for LLMs to directly make decisions on challenging technical problems only by prompting. To balance efficiency and interpretability, one approach is to integrate the LLMs into certain search strategies, where the genetic algorithm guides the search process and LLMs generate the candidate crossovers. LLMatic~\cite{nasir2023llmatic} and EvoPrompting~\cite{chen2023evoprompting} use code-LLMs as mutation and crossover operators for a genetic NAS algorithm. During evolution, each generation has a certain probability of deciding whether to perform crossover or mutation to produce new offspring. Crossover and mutation are generated by prompting LLMs. Such a solution integrates LLM into the genetic search algorithm, which would achieve better performances than direct reasoning.

The research mentioned above brings valuable insights to the field of automated learning in recommender systems. However, there are several challenges that need to be addressed. Firstly, the search space in recommender systems is considerably more complex, encompassing diverse types of search space and facing significant volume issues. This complexity poses a challenge in effectively exploring and optimizing the search space.
Secondly, compared to the common architecture search in other domains, recommender systems lack a strong foundation of knowledge regarding the informative components within the search space, especially the effective high-order feature interactions. Unlike well-established network structures in other areas, recommender systems operate in various domains and scenarios, resulting in diverse and domain-specific components. 
Addressing these challenges and advancing the understanding of the search space and informative components in recommender systems will pave the way for significant improvements in automated learning approaches.

\section{LLMs as Conversational Agent}
Conversational recommender system (CRS) is a specialized type of recommendation tool that aims to uncover users' interests and preferences through dialogue, enabling personalized recommendations and real-time adjustment of recommendation strategies based on user feedback. Compared to traditional recommender systems, conversational recommender systems have the advantage of real-time understanding of user intents and the ability to adapt recommendations based on user feedback. Typically, a conversational recommender system consists of two main components: a dialogue module and a recommendation module. 
In this section, we will primarily focus on discussing the dialogue module, which plays a crucial role in facilitating effective user-system interactions and understanding user preferences.

In a conversational recommender system, the dialogue module typically takes the form of a dialogue system. Dialogue systems can generally be classified into two main categories: chit-chat and task-oriented. The former focuses on open-domain question answering, and two major methods are commonly employed: generative and retrieval-based methods. Generative methods~\cite{shang2015neural,vinyals2015neural,sordoni2015neural} utilize a sequence-to-sequence model structure to generate responses, while retrieval-based methods~\cite{wu2019deep,qiu2015convolutional,wan2016deep} transform the task of generating responses into a retrieval problem by searching for the most relevant response in a response database based on the dialogue context. In conversational recommender systems, task-oriented dialogue systems are more often required, as they are specifically designed to assist users in accomplishing specific tasks.
For task-oriented dialogue systems, a common approach~\cite{sun2018conversational,greco2017converse} is to treat the response generation as a pipeline and handle it separately using four components: dialogue understanding~\cite{yao2013recurrent,mesnil2013investigation}, dialogue state tracking~\cite{goddeau1996form,henderson2013deep,mrkvsic2016neural}, dialogue policy learning~\cite{cuayahuitl2015strategic,sun2018conversational}, and natural language generation~\cite{zhou2016context,duvsek2016sequence}. Another approach is to employ an end-to-end method~\cite{wen2016network,bordes2016learning,zhang2019dialogpt}, training an encoder-decoder model to handle all the processing steps collectively. The first approach suffers from scalability issues and lacks synergy between the components, while the second approach requires a substantial amount of supervised data for training.

Based on the classification of dialogue systems, common approaches in conversational recommender systems can also be divided into two categories: attribute-based QA (question-answering) and generative methods. 
The attribute-based QA approach~\cite{sun2018conversational,lei2020estimation,lei2020interactive,deng2021unified} utilizes a pipeline method within the dialogue system. In each dialogue turn, the system needs to decide whether to ask the user a question or provide a recommendation. The decision-making process, particularly regarding which attribute to ask about, is typically handled by a policy network. On the other hand, generative methods do not explicitly model the decision-making process. Instead, they often employ an end-to-end training approach, where a sequence-to-sequence model generates output directly from a shared vocabulary of words and items. Whether the generated output is chit-chat, a question, or a recommendation is implicitly determined during the generation process.
Compared to attribute-based QA methods, generative methods~\cite{li2018towards,wang2022barcor,wang2022towards,wang2022recindial} appear to be simpler and more scalable. However, they require a large amount of supervised training data. With the advancement of pre-trained language models (PLMs) in the field of natural language processing, particularly models like BERT~\cite{devlin2018bert} and GPT~\cite{radford2018improving}, the capabilities of pre-trained models in language understanding and generation have become increasingly powerful. Researchers have found that fine-tuning pre-trained models with a small amount of supervised data can yield impressive results on specific tasks.
This discovery has led to the application of PLMs in generative conversational recommender systems. For example, DialoGPT~\cite{zhang2019dialogpt} achieved promising dialogue intelligence by fine-tuning GPT-2 on dialogue data collected from platforms like Reddit. Subsequently, BARCOR~\cite{wang2022barcor}, RecInDial~\cite{wang2022recindial}, and UniCRS~\cite{wang2022towards} utilized DialoGPT for constructing conversational recommender systems, with variations in their action decision strategies. While PLMs reduce the dependency of generative dialogue models on extensive data, the fine-tuning process still incurs significant computational time and requires the collection of high-quality domain-specific training data due to the large parameter space of the models.

With the increase in model parameters and training data, the intelligence and knowledge capacity of models continues to improve. OpenAI has been expanding the model parameters and training data while employing techniques such as RLHF (Reinforcement Learning from Human Feedback) and Instruction Tuning to further fine-tune GPT-3~\cite{brown2020language}. This has led to the emergent abilities of models like InstructGPT~\cite{ouyang2022training} and subsequent models like ChatGPT, which exhibit incredible intelligence and have opened the doors to new intelligent dialogue systems based on large language models (LLMs). Furthermore, Google's BARD and META's LLaMA~\cite{touvron2023llama} are also large language dialogue models that have been proposed and demonstrated remarkable performance in conversational abilities. The Vicuna model, for instance, utilizes dialogue corpora shared by users in using ChatGPT to fine-tune the open-source LLaMA model, with the team claiming it can achieve over 90\% of ChatGPT's capability. This series of successive LLM introductions has brought new insights to conversational recommender systems.
Due to the utilization of extensive open-domain corpora during LLM training, it possesses inherent conversational recommendation capabilities and can provide reasonable recommendations in open domains such as movies, music, and games. 

However, there are still significant \textbf{challenges} in building an enterprise-level CRS. 
The first challenge is the lack of awareness of large models about private domain data. It is well known that most of the training data for LLMs, such as GPT-3, comes from publicly available sources on the internet. As a result, these models may lack visibility into the data that resides within information platforms, making their modeling and understanding capabilities of such data relatively poor. To address this challenge, there are currently two approaches being explored: fine-tuning~\cite{zhang2019dialogpt} and tool learning~\cite{gao2023chat,friedman2023leveraging}. Fine-tuning involves tuning LLM using private domain-specific dialogue data. There are two major concerns in the approach. First, massive high-quality domain-specific dialogue data is required to tune the extremely large model. However, in most recommendation scenarios, data primarily consists of explicit or implicit user-item interactions, which may lack conversational context. Therefore, generating high-quality dialogue data from interaction data is a key concern in the approach. In RecLLM~\cite{friedman2023leveraging} and iEvaLM~\cite{wang2023rethinking}, researchers have proposed using LLMs to construct a user simulator for generating conversational data. Besides, the fine-tuning technique plays a crucial role in determining the ultimate quality of LLMs. A well-designed and effective fine-tuning strategy can lead to significant improvements in the model's performance and capabilities, such as instruction tuning and RLHF proposed in InstructGPT~\cite{brown2020language}. 
Tool learning is another approach to address this challenge, and its main idea is to treat traditional recommendation models as tools to be utilized, such as Matrix Factorization (MF) and DeepFM. For a more detailed explanation of tool learning, please refer to Section~\ref{tool_learning}. Since recommendation models are domain-specific, LLM can leverage these models to obtain recommendation results and recommend them to the users in the response. In this approach, there are two main technical points: the construction of the tool model and the engineering of prompts to guide the LLM in the proper utilization of the tool. First of all, conventional recommendation models generally use id or categorical features as input, while users always give their requirements or preferences in natural language in conversations. Therefore, unstructured text features should be taken into consideration in tool construction. In Chat-Rec~\cite{gao2023chat}, a conventional recommendation model and a text embedding-based model(text-embedding-ada-002) are used as tools. RecLLM~\cite{friedman2023leveraging} adapted a language model enhanced dual-encoder model and several text retrieval methods as the recommendation engine. On the other hand, despite the strong intelligence and reasoning capabilities of LLMs, effectively harnessing these abilities requires well-crafted prompts for guidance. For instance, the Chain of Thought proposed by Jason~\cite{wei2022chain} could trigger LLM to reason and engage in step-by-step thinking, which benefits the tool-using capability. Subsequent studies like ToT~\cite{yao2023tree}, Plan-and-Solve~\cite{wang2023plan} and ReAct~\cite{yao2022react} have proposed more advanced techniques for prompt design to assist in guiding LLM to engage in deeper thinking and tool planning. 

The second challenge lies in the issue of memory and comprehension in long conversations. Due to the input constraints of LLMs, models like ChatGPT can support a maximum of 4096 tokens in a single call, including both input and output. In multi-turn dialogue scenarios, longer dialogue contexts often meet the risk of exceeding this token limit. The simplest approach to tackle this challenge is to trim the dialogue by discarding earlier turns. However, in conversational recommender systems, users may express a significant amount of personal information and interests in the early stages of the conversation. The omission of such information directly impacts the accuracy of recommendations. To address this issue, several relevant works have proposed solutions. MemPrompt~\cite{madaan2022memory} enhances the prompt by incorporating a memory module, enabling GPT-3 to possess stronger long-dialogue memory capability. Similarly, RecLLM~\cite{friedman2023leveraging} leverages LLM to extract user profiles and store them as factual statements in user memory. When processing user queries, relevant facts are retrieved based on text similarity.  

\section{Tool-Learning and its Applications in Recommendation} \label{tool_learning}

\subsection{LLM-based Tool Learning}
Tool learning is an emerging research field that aims to enhance task-solving capabilities by combining specialized tools with foundational models, which has been understood by~\cite{qin2023tool} as two perspectives: 
\begin{enumerate}
    \item \textbf{Tool-augmented learning} treats specialized tools as assistants in order for improving the quality and accuracy of tasks, or \textbf{Tool for AI};
    \item \textbf{Tool-oriented learning} focuses more on training models to effectively use tools, controlling and optimizing tool-applying processes, or \textbf{AI for Tool}.
\end{enumerate}

Tool learning has found applications in various fields, and this section primarily focuses on tool learning paradigms based on large language models (LLMs). While recent works often involve a combination of these two perspectives, we do not specifically categorize each work into one type. LLMs, such as GPT, are well-suited for tool learning applications~\cite{mialon2023augmented}. With their powerful natural language processing capabilities, LLMs can break down complex tasks into smaller sub-tasks and convert them into executable instructions. Specialized tools allow LLMs to access knowledge that is beyond their own understanding. By integrating specialized tools, LLMs can better understand and address complex problems, offering more accurate and efficient solutions.

\begin{table*}[]
\newcommand{\tabincell}[2]{
\begin{tabular}{@{}#1@{}}#2\end{tabular}
}

    \caption{LLM-based tool learning approaches}
    \label{tab:toollearning}
    \centering
    \resizebox{1.0\textwidth}{!}{
\begin{tabular}{c|c|c|c}
\toprule
Approach & Tool usage & LLM backbone & Task \\ \midrule
Re3\cite{yang2022re3} & LLM & \tabincell{c}{gpt3-instruct-175B\\ gpt3-instruct-13B} & Long Stories Generation\\ \midrule
PEER\cite{schick2022peer} & LLM & LM-Adapted T5 & Editions, Citations, Quotes\\ \midrule
METALM\cite{hao2022language} & \tabincell{c}{Pretrained Encoders \\ with diverse modalities} & \tabincell{c}{Transformer \\ (pretrained from scratch)} & \tabincell{c}{language-only tasks\\ vision-language tasks}\\ \midrule
Atlas\cite{izacard2022few} & Dense retriever & T5 & \tabincell{c}{Knowledge-Intensive Language Tasks\\ Massively-Multitask Language Understanding\\ Question Answering\\ Fact Checking}\\ \midrule
LaMDA\cite{thoppilan2022lamda} & \tabincell{c}{Retriever\\ Translator\\ Calculator} & Decoder-only Transformer & Dialog\\ \midrule
WebGPT\cite{nakano2021webgpt} & Web Browser & gpt-3 & Question answering\\ \midrule
Mind's Eye\cite{liu2022mind} & \tabincell{c}{Physics Engine\\ Text-to-code LM} & \tabincell{c}{gpt-3\\ PaLM} & Reasoning\\ \midrule
PAL\cite{gao2022pal} & Python Interpreter & CODEX(code-davinci-002) & \tabincell{c}{Mathematical\\ Symbolic\\ Algorithmic Reasoning}\\ \midrule
SayCan\cite{ahn2022can} & Robots & PaLM & Real-world robotic tasks\\ \midrule
HuggingGPT\cite{shen2023hugginggpt} & \tabincell{c}{AI models\\ in Hugging Face Community} & \tabincell{c}{gpt-3.5-turbo\\ text-davinci-003 \\ gpt-4} & \tabincell{c}{Image Classification\\ Image Captioning\\ Object Detection\\ etc.}\\ \midrule
Auto-GPT & \tabincell{c}{Web Browser} & \tabincell{c}{gpt-3.5-turbo\\text-davinci-003\\ gpt-4} & User-specified Tasks\\ \midrule
\tabincell{c}{Visual ChatGPT\cite{wu2023visual} \\ Taskmatrix.AI\cite{liang2023taskmatrix}} & \tabincell{c}{Visual Foundation models\\ Customized models with unified API form} & text-davinci-003 & Visual Customized Tasks\\\midrule
ReAct\cite{yao2022react} & Wikipedia API & PaLM-540B & \tabincell{c}{Question Answering\\ Face Verificaiton}\\ \midrule
Toolformer\cite{schick2023toolformer} & \tabincell{c}{Calculator\\ Q\&A system\\ Search Engine \\ Translation System\\ Calendar} & GPT-J & Downstream Tasks\\
 \bottomrule                                   
\end{tabular}
}
\end{table*}

LLMs are commonly applied as controllers to select and manage various existing AI models to solve complex tasks, which rely on user input and language interfaces on making summarizations. They act as the central component, responsible for comprehending problem statements and making decisions regarding which actions to execute. Additionally, they aggregate the outcomes based on the results of the executed actions. In that case, HuggingGPT~\cite{shen2023hugginggpt} leverages existing models from the Hugging Face community\footnote{https://huggingface.co} to assist in task-solving. Visual ChatGPT\cite{wu2023visual} combines visual foundation models like BLIP~\cite{li2022blip}, Stable Diffusion~\cite{rombach2022high}, etc. with LangChain\footnote{https://docs.langchain.com} to handle complex visual tasks, while the following TaskMatrix.AI~\cite{liang2023taskmatrix} maintains a unified API Platform extending the capabilities of Visual ChatGPT, extends the capabilities of Visual ChatGPT by maintaining a unified API Platform, enabling input from multiple modalities and generating more complex task solutions. On the contrary, Auto-GPT\footnote{https://github.com/Significant-Gravitas/Auto-GPT} operates as an agent that autonomously understands specific targets through natural language and performs all processes in an automated loop, without requiring mandatory human input.
WebGPT~\cite{nakano2021webgpt} introduces a text-based web browsing interactive environment, where LLMs learn to emulate the complete process of human interaction with a web browser using behavior cloning and rejection sampling techniques. In ReAct~\cite{yao2022react}, by leveraging an intuitive prompt, LLMs learn to generate both reasoning paths and task-specific actions alternately when solving a specific task. The execution of specific actions is delegated to corresponding tools, and external feedback obtained from these tools is utilized to validate and guide the reasoning process further. The motivation behind Toolformer~\cite{schick2023toolformer} aligns closely with ReAct; however, it goes a step further by combining diverse tools within a single model. This integration provides the model with flexible decision-making abilities and improved generalization capabilities, achieved through a simple yet effective self-supervised method. In contrast to prior works, LATM~\cite{cai2023large} takes a novel approach by empowering LLMs to directly generate tools. It achieves a division of labor within the task-solving process by employing LLMs at different scales: the tool maker, tool user, and dispatcher. LATM is entirely composed of LLMs, enabling the self-generation and self-utilization of tools.

\subsection{Applications in Personalization Scenarios}
Recently, LLMs have demonstrated impressive abilities in leveraging internal world knowledge and common sense reasoning to accurately understand user intent from dialogues. Moreover, LLMs can communicate with users fluently in natural language, offering a seamless and delightful user experience. These advantages make LLMs an appealing choice as recommendation agents to enhance the personalized experience.

However, despite the impressive memory capacity of LLMs, they face challenges in memorizing specific knowledge in private and specialized domains without sufficient training. For instance, storing the item corpus and all user profiles in a recommender system can be challenging for LLMs. This limitation can result in LLMs generating inaccurate or incorrect responses and makes it difficult to control their behavior within a specific domain. Furthermore, LLMs face the challenge of the \emph{temporal generalization problem} as external knowledge continues to evolve and change over time. To address these issues, various tools can be utilized to augment LLMs and enhance their effectiveness as recommendation agents.

\textbf{Search engine.} Search engines are widely employed to provide external knowledge to LLMs, reducing LLMs’ memory burden and alleviating the occurrence of hallucinations in LLMs’ responses. BlenderBot 3~\cite{shuster2022blenderbot} uses specific datasets to fine-tune a series of modules, enabling LLMs to learn to invoke the search engine at the appropriate time and extract useful knowledge from the retrieval results. LaMDA~\cite{thoppilan2022lamda} learns to use a toolset that includes an IR system, a translator, and a calculator through fine-tuning to generate more factual responses. RETA-LLM~\cite{liu2023reta} is a toolkit for retrieval-augmented LLMs. It disentangles IR systems and LLMs entirely, facilitating the development of in-domain LLM-based systems.~\cite{thoppilan2022lamda} shows a case of applying LaMDA to content recommendation. Preconditioned on a few role-specific dialogues, LaMDA can play the role of a music recommendation agent.

\textbf{Recommendation engine.} Some works have attempted to alleviate the memory burden of LLMs by equipping them with a recommendation engine as a tool, enabling LLMs to offer recommendations grounded on the item corpus. The recommendation engine in Chat-REC\cite{gao2023chat} is further divided into two stages: retrieve and reranking, which aligns with typical recommendation system strategies. In the retrieval stage, LLMs utilize traditional recommendation systems as tools to retrieve 20 items from the item corpus as a candidate item set. Subsequently, LLMs employ themselves as tools to rerank the candidate item set. LLMs' commonsense reasoning ability, coupled with the internal world knowledge within them, allow them to provide explanations for the sorting results. The recommendation engine tool used in RecLLM\cite{friedman2023leveraging} is highly similar to it in Chat-REC,  and it is also divided into retrieval and reranking stages. RecLLM provides several practical solutions for large-scale retrievals, such as Generalized Dual Encoder Model and Concept Based Search, and so on.

\textbf{Database.} Databases are also utilized as tools to supplement additional information for LLMs. In order to better cope with the cold-start problem for new items and alleviate the temporal generalization problem of LLMs, a vector database is utilized to provide information for new items that the LLMs are unaware of in Chat-REC~\cite{gao2023chat}. When encountering new items, LLMs can utilize this database to access information about them based on the similarity between the user’s request embedding and item embeddings in the database. User profiles can also help LLMs better understand the user's intent. RecLLM\cite{friedman2023leveraging} employs a user profile module as a tool to deposit meaningful and enduring facts about users exposed during historical conversations in user memory and retrieve a single fact related to the current dialogue when necessary.

Although some works have applied the concept of tool learning to personalization systems, there are still interesting and promising research topics that deserve exploration. 
1) \textbf{Fine-tuning models for better tool use.} In-context learning has shown promise in teaching LLMs how to effectively use tools with a small number of demonstrations, as shown in Chat-REC and RecLLM. However, LLMs often struggled to learn strategies for handling complex contexts with limited demonstrations. Fine-tuning is a viable option for improving tool use, but it requires sufficient training data and effective techniques. RecLLM further fine-tunes some modules of it using synthetic data generated by a user simulator through RLHF\cite{ouyang2022training} technique. Investigating methods to obtain sufficient training data and developing tailored fine-tuning techniques for recommendation systems is a worthwhile research direction. 
2) \textbf{Developing a more powerful recommendation engine.} Traditional recommendation systems often rely on collaborative filtering signals and item-to-item transition relationships for recommendations. However, with the use of LLMs as the foundation models, user preferences can be reflected through natural language and even images. Therefore, developing a recommendation engine that supports multimodal data is a crucial research direction. Additionally, the recommendation engine should also be capable of adjusting the candidate set based on user preferences or feedback (such as querying movies of a specific genre or disliking an item in the recommendation set).  
3) \textbf{Building more tools.} To provide LLMs with more authentic and personalized information, the development of additional tools is crucial. For example, APIs for querying knowledge graphs~\cite{pan2023unifying} or accessing users' social relationships can enhance the knowledge available to LLMs, enabling more accurate and tailored recommendations.

\section{LLMs as Personalized Content Creator}\label{sec:llm_personalized_content_creator}

Traditional recommender systems focus on suggesting existing items based on user preferences and historical data, where displayed content is already generated for retrieval. However, with the advancements in techniques and platforms for content creators, personalized content creator has attracted more and more attention, where more appealing content is customized generated to match the user's interests and preferences, especially in the realm of online advertising~\cite{vempati2020enabling}. The common contents contain the visual and semantic contents~\cite{thomaidou2013automated,zhang2022automatic,lei2022plato}, such as title, abstract, description, copywritings, ad banners, thumbnail, and videos. One more widely discussed topic is text ad generation, where the ad title and ad description are generated with personalized information. Earlier works adopt the pre-defined templates~\cite{bartz2008natural,fujita2010automatic,thomaidou2013automated} to reduce the extensive human effort, which, however, often fail to fully meet the user's interests and preferences. More recent data-driven methods have emerged, which incorporate user feedback as rewards in the reinforcement learning framework to guide the generation process~\cite{hughes2019generating,wang2021reinforcing,chen2022personalized,zhang2021chase}. Furthermore, the incorporation of pre-trained language models has played a significant role in improving the generation process for multiple content items~\cite{kanungo2021ad,wei2022creater,kanungo2022cobart,chen2019towards}. This integration helps refine the content generation models and improve their ability to meet user preferences effectively.

As recommender systems and large language models continue to evolve, a promising technique that would bring new opportunities is the integration of AI Generated Content (AIGC). AIGC~\cite{cao2023comprehensive} involves the creation of digital content, such as images, music and natural language through AI models, with the aim of making the content creation process more efficient and accessible.  Earlier efforts in this field focused on deep-learning-based generative models, including Generative Adversarial Networks (GANs)~\cite{goodfellow2014generative}, Variational AutoEncoders (VAEs)~\cite{kingma2013auto}, Normalizing Flows~\cite{dinh2014nice}, and diffusion-based models~\cite{ho2020denoising} for high-quality image generation. As the generative model evolves, it eventually emerges as the transformer architecture~\cite{vaswani2017attention}, acting as the foundational blocks for BERT~\cite{kenton2019bert} and GPT~\cite{radford2018improving} in the field of NLP, and for Vision Transformer (ViT)~\cite{dosovitskiyimage} and Swin Transformer~\cite{liu2021swin} in the field of CV. Moreover,  the scope of generation tasks expanded from uni-modal to multi-modal tasks, including the representative model CLIP~\cite{radford2021learning}, which can be used as image encoders with multi-modal prompting for generation. The multi-modal generation has become an essential aspect of AIGC, which learns the multimodal connection and interaction, typically including vision language generation~\cite{radford2021learning}, text audio generation~\cite{chenadaspeech}, text graph generation~\cite{li2016commonsense}, text Code Generation~\cite{feng2020codebert}. With the emergence of large language models, nowadays AIGC is achieved by extracting the human intention from instructions and generating the content according to its knowledge and intention. Representative products, including ChatGPT~\cite{chatgpt}, DALL-E-2~\cite{ramesh2021zero}, Codex~\cite{chen2021evaluating} and Midjourney~\cite{midjourney}, have attaining significant attention from society. With the growth of data and model size, the model can learn more comprehensive information and thus leading to more realistic and high-quality content creators.

Recall to the personalized content creator, the large language models would bring opportunities from the following points. Large language models would further extend the capabilities of the pre-trained model, allowing for better reasoning of user personalized intent and interest. Previous methods~\cite{wei2022creater,kanungo2022cobart} depending on tailored pre-training models may be enhanced to better improve the reasoning abilities and few-shot prompting. Secondly, \textit{Reinforcement Learning from Human Feedback} (RLHF) strategy can be applied to fine-tune models to better capture the user intent information, similar to existing RL-based framework~\cite{wang2021reinforcing} for text ad generation. Last but not least, the powerful generative abilities of large language models empower realistic creation thanks to the availability of sufficient cross-modal knowledge bases. The work~\cite{wang2023generative} more specifically proposes a recommendation paradigm based on ChatGPT, where the generation process receives feedback and multiple rounds of conversions to better capture the user explicit preferences. Compared to previous training paradigms, more explicit expressions of user interest can be understood by the large language models and converted into corresponding instructions to guide the generation of content, significantly alleviating the problem of extremely sparse feedback.

However, there are two major security and privacy risks for personalized content creators. One of the concerns is the reliability of models like ChatGPT in terms of factuality, as indicated in the work ~\cite{borji2023categorical}. While these models generate content that appears reasonable, there is a risk of distributing misleading or inaccurate information, which can weaken the truthfulness of internet content. This concern becomes particularly crucial in personalized recommendations, where the model may inadvertently promote misleading information tailored to the user's interests.
The second concern revolves around data privacy, encompassing both user profiles and long-term human interaction histories. In the case of large language models, these interaction histories are collected or shared, potentially leading to the large models memorizing sensitive user data. Previous work~\cite{carlini2021extracting} has demonstrated that large language models, especially GPT-2~\cite{radford2019language}, memorize and leak individual training examples. This emphasizes the need for strict user approval and careful handling of annotator data to mitigate privacy risks. It is crucial to develop new techniques that prioritize privacy preservation during the training process.

\section{Open Challenges}\label{sec:challenges}

\subsection{Industrial Challenges} Personalization services, particularly with recommender systems, are complex industrial products that face numerous challenges when implemented in real-world scenarios. We will now summarize the key challenges as follows:

    \textbf{Scaling computational resources} Existing large language models, such as BERT and GPT, demand significant computational power for training and inference. This includes high memory usage and time consumption. Fine-tuning these models to align them with personalization systems, which has shown promising results for improved personalization performance, can be computationally intensive. Several efficient finetuning strategies, e.g., option tuning in M6-Rec~\cite{cui2022m6}, Lora~\cite{hu2021lora}, QLora~\cite{dettmers2023qlora}, have been developed to address this issue and pave the way for more efficient tuning.

    \textbf{Significant Response time} Achieving efficient response times is crucial for online serving and greatly impacts the personalized user experience. Response time includes both the inference phase of large language models and the concurrent user requests in large numbers. The introduction of large language models can result in considerable inference time, posing a challenge for real-world deployment. One approach is to pre-compute the embeddings of intermediate outputs from language models, storing and indexing them in a vector database, particularly for methods that utilize large language models as textual encoders. Other approaches, such as distillation and quantization, aim to strike a balance between performance and latency.

\subsection{Laborious Data Collection} Large language models are widely known to leverage extensive amounts of open-domain knowledge during their training and fine-tuning processes. These knowledge sources include well-known references such as Wikipedia, books, and various websites~\cite{brown2020language}. Similarly, when applied in recommender systems, these models often rely on representative open-domain datasets such as MovieLens and Amazon Books. While this type of open-domain knowledge contains a wealth of common-sense information, personalized tasks require access to more domain-specific data that is not easily shareable. Additionally, the nature of user feedback in personalized tasks can be complex and sparse, often accompanied by noisy feedback. Collecting and filtering this data, in contrast to acquiring common-sense knowledge, presents challenges. It incurs higher labor costs and introduces additional training redundancy due to the need for extensive data processing and filtering. Furthermore, designing appropriate prompts to instruct or fine-tune large language models is crucial for aligning them with the distribution of in-domain inputs in personalization tasks. By carefully tailoring the prompts, researchers and practitioners can guide the model to produce outputs that better cater to personalized applications, thereby maximizing performance and effectiveness.

\subsection{Long Text Modeling}
Large language models have a limitation on the maximum number of input tokens they can handle, typically constrained by the context window size, e.g., 4096 for ChatGPT. This poses challenges when dealing with long user behavior sequences, which are common in modern recommender systems. Careful design is necessary to generate effective and appropriate prompt inputs within this limited length. In the case of conversations with multiple rounds, accumulating several rounds of dialogue can easily exceed the token limit of models. The current approach in handling long conversations is to truncate the history, keeping only the most recent tokens. However, this truncation discards valuable historical information, potentially harming the model performance. To address these challenges, several techniques can be employed. One approach is to prioritize and select the most relevant parts of the user behavior sequence or conversation history to include in the prompt. This selection can be based on various criteria such as recency, importance, or relevance to the task at hand. Another technique involves summarizing or compressing the lengthy input while preserving essential information. This can be achieved through techniques like extractive summarization or representing the long sequence in a condensed form. Moreover, architectural modifications, such as hierarchical or memory-augmented models, can be explored to better handle long sequences by incorporating mechanisms to store and retrieve relevant information efficiently. 

In addition, collaborative modeling of long text data and recommendation tasks is an emerging and pressing challenge. In conventional personalization systems, item ID information along with other categorical information is commonly used for modeling feature interactions and user preferences. With the rise of large language models, there would be a growing trend toward leveraging textual information more extensively. Textual data provides unique insights about items or users, making it valuable for modeling purposes. From the perspective of modeling, dealing with long text data requires more attention and complexity compared to categorical data, not to mention the need to match the modeling of user interests. From the perspective of implementation, reforming the entire pipeline becomes necessary to accommodate the requirements of efficient latency. Efficiently processing and incorporating long text data into recommendation models and serving them in real-time present technical challenges.

\subsection{Interpretability and Explainability}
While large language models provide good reasoning capabilities, they are notorious for the nature of the 'black box', which is highly complex and non-linear in their enormous size and layered architecture, making it challenging to comprehend the internal workings and understand the generation process of recommendations. Without a deep understanding of how the model operates, it becomes challenging to detect and address biases or ensure fair and ethical recommendations. Once transparency about the internal mechanisms is lacking, users struggle to trust and accept the decisions made by the system. Users often desire understandable explanations for recommended choices. Addressing the challenge of model interpretability and explainability requires research involving natural language processing, explainable AI, human-computer interaction, and recommendation systems. The development of techniques that unveil the inner workings of language models, facilitate the generation of meaningful and accurate interpretations, and enable robust evaluation methods is the main focus. By providing transparent and interpretable recommendations, users can establish trust, understand the reasoning behind the recommendations, and make informed decisions.

\subsection{Evaluation}
Conventional personalization systems typically rely on task-specific metrics such as ranking-oriented metrics, NDCG, AUC, and Recall to evaluate model performance. However, with the integration of large language models into recommender systems, the evaluation tools and metrics undergo significant changes. Traditional metrics may not sufficiently capture the performance of recommender systems powered by large language models, which introduce novel capabilities and generate recommendations in a different manner and require the development of new evaluation tools. 

One crucial aspect of evaluation is considering user preferences in large language model-powered systems, which requires a user-centric approach. Metrics such as user satisfaction, engagement, and overall experience become essential considerations. For example, Liu's work~\cite{liu2023chatgpt} proposes a crowdsourcing task to assess the quality of generated explanations and review summaries, providing a way to evaluate the effectiveness of the generated content. Additionally, user satisfaction surveys and feedback questionnaires can serve as valuable options.

Another perspective to consider is the health of the system, which involves evaluating novelty and assessing factors like diversity, novelty, serendipity, and user retention rates. These metrics help evaluate the freshness of recommendations and the long-term effects of large language models.

Furthermore, it is crucial to assess the interpretability and fairness of recommendations. The interpretability assessment focuses on measuring the clarity, understandability, and transparency of recommendations. Simultaneously, the fairness evaluation aims to address potential biases in personalized results. By prioritizing fairness, we strive to create personalized experiences that are equitable and inclusive for all users. Both of these evaluations are essential to enhance the overall user experience and build confidence in the personalized recommendations delivered by the system.

\subsection{Trade-off between Helpfulness, Honesty, Harmlessness}
When large language models are employed for personalization, some of their disadvantages would be magnified. Striving for a more honest and harmless system may come at the expense of system performance.

First of all, the accuracy and factuality of the system must be ensured. Although large language models can generate seemingly reasonable content, there is a risk of disseminating misleading or inaccurate information. This becomes even more critical when incorporating user feedback, as the model may mimic user behaviors in an attempt to appear honest. However, this imitation can result in biased guidance for users, offering no real benefits.

Secondly, in terms of harmlessness, concerns regarding privacy, discrimination, and ethics arise. While large language models have the potential to provide highly personalized recommendations by leveraging user data, privacy, and data security become paramount. Unlike open-domain datasets, the privacy of individual data used for training should be rigorously protected, with strict user permissions for sharing their personal information. For discrimination, large language models may inevitably reflect biases inherent in the training data, leading to discriminatory recommendations. Considering the biased user and item distribution, which is much more significant in recommender systems with the long-tail effect, where biased user and item distribution can lead to decisions that favor majority choices, resulting in discrimination against certain users. The final concern revolves around ethical considerations. Harmful messages, if clicked by users unconsciously, can guide large language models toward generating similar harmful content. However, when assisting in personalized decision-making, it is essential for large language models to have the capability to minimize exposure to harmful messages and guide users in a responsible manner. Approaches like constructing a Constitutional AI~\cite{bai2022constitutional}, where critiques, revisions, and supervised Learning are adopted for better training large language models, may offer valuable insights.

By addressing these concerns, safeguarding privacy, mitigating discrimination, and adhering to ethical guidelines, recommender systems can leverage the power of large language models while ensuring user trust, fairness, and responsible recommendations.

\section{Conclusion}
In conclusion, the emergence of large language models represents a significant breakthrough in the field of artificial intelligence. Their enhanced abilities in understanding, language analysis, and common-sense reasoning have opened up new possibilities for personalization. In this paper, we provide several perspectives on when large language models adapt to personalization systems. We have observed a progression from utilizing low-level capabilities of large language models to enhance performance, to leveraging their potential in complex interactions with external tools for end-to-end tasks. This evolution promises to revolutionize the way personalized services are delivered. We also acknowledge the open challenges that come with the integration of large language models into personalization systems.

\bibliographystyle{IEEEtran}
\bibliography{ref}

\end{document}